\documentclass[a4paper,11pt]{article}

\usepackage{amsmath,amssymb,mathtools}
\usepackage{braket}
\usepackage{color}
\usepackage{graphicx}
\usepackage{subfigure}
\usepackage{cite}
\usepackage[colorlinks=true,linkcolor=blue, citecolor=red, urlcolor=blue, bookmarks]{hyperref}
\usepackage{multirow,makecell} 
\usepackage[figuresright]{rotating} 
\usepackage{textcomp}
\usepackage{wasysym}
\usepackage{ulem}

\usepackage[utf8]{inputenc}
\usepackage[T1]{fontenc}

\usepackage[text={17.1cm,24.6cm},centering]{geometry} 

\numberwithin{equation}{section}

\def \be {\begin{equation}}
\def \ee {\end{equation}}
\def \ba {\begin{array}}
\def \ea {\end{array}}
\def \bea{\begin{eqnarray}}
\def \eea{\end{eqnarray}}

\def \ve {\varepsilon}

\def \l {\lambda}
\def \lam {\lambda}

\def \s {\sigma}

\def \r {\rho}

\def \vph {\varphi}

\def \f {\frac}

\def \inf {\infty}

\def \lag {\langle}
\def \rag {\rangle}

\def \ep {\mathrm{e}}
\def \ii {\mathrm{i}}

\def \tr {\textrm{tr}}

\def \and {{~\textrm{and}~}}

\def \GGE {{\textrm{GGE}}}

\begin{document}

\title{\textbf{
Discrete power-law decay of subsystem distance after a quantum quench
}
}
\author{
Bin Sui
and
Jiaju Zhang\footnote{Corresponding author: jiajuzhang@tju.edu.cn}
}
\date{}
\maketitle
\vspace{-10mm}
\begin{center}
{\it
Center for Joint Quantum Studies and Department of Physics, School of Science, Tianjin University,\\
135 Yaguan Road, Tianjin 300350, China
}
\vspace{10mm}
\end{center}

\begin{abstract}

  We present a numerical study of subsystem distance decay following a global quantum quench in the infinite one-dimensional transverse-field Ising chain, using the mathematically rigorous Bures distance  $B_A(t)$ to quantify the deviation of the time-evolved reduced density matrix from its stationary generalized Gibbs ensemble state. We show that the late-time decay follows a discrete power law $B_A(t) \sim t^{-\lambda}$, with the exponent $\lambda$ confined to discrete values: $1$, $5/4$, $3/2$, $7/4$, $2$, $5/2$, and potentially further values. The specific exponent is jointly determined by the pre- and post-quench transverse fields, as well as by properties of the symmetric excitation-fraction function $m_S(\varphi)$, defined on $\varphi\in[0,\pi]$ to characterize the pre-quench Hamiltonian eigenstates, including continuity, boundary values, and first-derivative boundary values, among others. The previously established $t^{-3/2}$ decay for the initial ground state of the pre-quench Hamiltonian is naturally recovered as a special case of this general classification. Our results reveal a universal discrete structure governing local equilibration dynamics in integrable quantum systems.

\end{abstract}

\maketitle

\newpage

\tableofcontents


\section{Introduction}

The non-equilibrium dynamics of quantum many-body systems stands as one of the most fundamental and active frontiers in modern condensed matter physics and quantum statistical mechanics \cite{Polkovnikov:2010yn,Eisert:2014jea,Santos:2018rwj,Stefanucci:2025qmh}. Driven by rapid experimental advances in ultracold atomic gases, trapped ion systems and superconducting qubits \cite{Toshiya:2004qcv,Kinoshita:2006xby,Bloch:2012uep,Daley:2012xhf,Islam:2015mom}, far-from-equilibrium states of quantum matter can now be prepared and probed with high precision, motivating extensive theoretical efforts to uncover the universal principles governing relaxation, thermalization and entanglement spreading in closed quantum systems.

A core question is whether and how an isolated quantum system initialized in a non-equilibrium state relaxes to a stationary state, and what statistical ensemble describes its long-time behavior. For generic chaotic many-body systems, the eigenstate thermalization hypothesis (ETH) \cite{Deutsch:1991msp,Srednicki:1994mfb,Rigol:2007mja} provides the canonical framework: individual energy eigenstates reproduce thermal expectation values for local observables, such that the system relaxes to an ordinary thermal equilibrium state described by the canonical or microcanonical ensemble. In sharp contrast, integrable systems host an extensive set of conserved quantities that forbid conventional thermalization. Instead, their long-time stationary states are described by the generalized Gibbs ensemble (GGE) \cite{Rigol:2006jrd,Vidmar:2016laa}, which incorporates all local and quasi-local conservation laws and correctly captures steady-state values of local observables and correlation functions \cite{Calabrese:2011vdk,Calabrese:2012fmm,Calabrese:2012lgi}.

The global quantum quench, where a system is prepared in an eigenstate of a pre-quench Hamiltonian and then evolved unitarily under a suddenly tuned post-quench Hamiltonian, has become the standard paradigm for investigating non-equilibrium quantum dynamics. Over the past decades, extensive studies on quantum quenches have revealed a wealth of universal dynamical phenomena, including the light-cone propagation of correlations \cite{Cheneau:2012zdz}, the ballistic growth of entanglement entropy \cite{Calabrese:2005in}, and the power-law relaxation of order parameters and correlation functions \cite{Calabrese:2005in}. The one-dimensional transverse-field Ising chain, which is exactly solvable via Jordan-Wigner, Fourier and Bogoliubov transformations \cite{Lieb:1961fr,Katsura:1962hqz,Pfeuty:1970ayt}, has served as a prototypical platform for such studies, enabling precise characterization of post-quench dynamics for both ground and excited initial states \cite{Bucciantini:2014urw}.

While the stationary properties of post-quench states are by now well established, the dynamical process through which a local subsystem approaches its steady-state reduced density matrix (RDM) remains less fully explored. Quantifying the distance between the time-dependent subsystem RDM and its GGE counterpart provides direct insight into the microscopic mechanism of local equilibration and the characteristic timescales of relaxation. In a pioneering work \cite{Fagotti:2013jzu}, Fagotti and Essler investigated the decay of a subsystem RDM distance measure after a quantum quench in the transverse-field Ising chain, and found that for initial ground states the distance follows a $t^{-3/2}$ power-law decay at late times. Despite its importance, this study suffers from three limitations. First, it is restricted exclusively to initial ground states of the pre-quench Hamiltonian, leaving the relaxation behavior of general excited eigenstates largely unaddressed. Second, it does not systematically survey the full parameter space of pre- and post-quench transverse fields, including quenches from or to quantum critical points. Third, the distance measure adopted there, constructed from the trace of the normalized squared difference of density matrices, has not been proven to satisfy the axioms of a mathematically rigorous distance metric on the space of quantum states.

In this paper, we overcome all three limitations above and present a comprehensive numerical study of the late-time decay of subsystem distance after a global quantum quench in the infinite transverse-field Ising chain. To remedy the lack of mathematical rigor in the distance measure of prior work, we adopt the Bures distance, a canonical metric rigorously defined on the space of density operators. Its construction is rooted in the Uhlmann fidelity, which quantifies the state overlap between two arbitrary density matrices and is formally defined as \cite{Nielsen:2010oan}
\be
F(\rho,\sigma) \equiv \tr\sqrt{\sqrt{\rho}\,\sigma\,\sqrt{\rho}}.
\ee
The Bures distance is then derived from the fidelity according to
\be
B(\rho,\sigma) \equiv \sqrt{2(1 - F(\rho,\sigma))},
\ee
which satisfies all axioms of a metric: non-negativity, symmetry to its inputs, identity of indiscernibles, and the triangle inequality. The subsystem Bures distance $B_A(t) \equiv B(\r_A(t),\r_{A,GGE})$ thus provides a mathematically consistent and operationally well-founded measure of the discrepancy between the time-evolved subsystem RDM $\r_A(t)$ and the GGE RDM $\r_{A,\GGE}$. Making use of a recently developed numerical algorithm for computing the Bures distance between fermionic Gaussian states \cite{Guo:2025jsm}, we systematically explore a broad range of quench parameter pairs $(h_0, h)$ covering paramagnetic, ferromagnetic and critical regimes. We consider general Gaussian initial eigenstates characterized by a momentum-dependent excitation fraction $m(\varphi)$ \cite{Alba:2009th,Bucciantini:2014urw}, and demonstrate that the late-time decay of the Bures distance universally follows a power law $B_A(t) \sim t^{-\lambda}$ where the decay exponent $\lambda$ takes only discrete values: $1$, $5/4$, $3/2$, $7/4$, $2$, $5/2$, and potentially further values. We show that the specific value of $\lambda$ is determined jointly by the pre- and post-quench transverse fields and the properties of the symmetric combination $m_S(\varphi) \equiv m(\varphi) + m(-\varphi) - 1$ with $\varphi \in [0,\pi]$, including its continuity, boundary values and boundary values of its first-derivative, and possible other properties. The previously reported $t^{-3/2}$ decay for ground-state initial conditions is recovered as a special case of our general classification.

The remainder of this paper is organized as follows.
In Sec.~\ref{sec_pro}, we introduce the transverse-field Ising model and the quantum quench protocol.
In Sec.~\ref{sec_typ}, we identify five representative sets of pre- and post-quench transverse field pairs that capture all qualitatively distinct relaxation regimes.
In Sec.~\ref{sec_ide}, we present our main results on discrete power-law decay exponents, and provide a classification scheme based on the properties of $m_S(\varphi)$.
Finally, we summarize our findings and discuss open questions in Sec.~\ref{sec_con}.

\section{Quench protocol} \label{sec_pro}

We consider a transverse-field Ising chain of $L$ sites, described by the Hamiltonian
\be \label{Hh}
H(h) = - \f{1}{2} \sum_{j=1}^{L} ( \sigma_j^x \sigma_{j+1}^x + h \sigma_j^z ),
\ee
where Pauli matrices $\sigma_j^\mu$ with $\mu=x,y,z$ act on the $j$-th site, and periodic boundary conditions are imposed such that $\sigma_{L+1}^x=\sigma_1^x$. The model is exactly solvable via successive Jordan-Wigner, Fourier, and Bogoliubov transformations \cite{Lieb:1961fr,Katsura:1962hqz,Pfeuty:1970ayt}. In the thermodynamic limit $L\to\inf$, the single-particle energy-momentum dispersion relation reads
\be
\ve(\vph) = \sqrt{ h^2 - 2h \cos\vph +1 }, ~ \vph\in(-\pi,\pi].
\ee
The model hosts a quantum phase transition at the critical point $|h|=1$: for $|h|<1$, the ground state lies in the ferromagnetic phase, while for $|h|>1$ it resides in the paramagnetic phase. As shown in Figure~\ref{figh0h}, over the full range $h\in(-\inf,\inf)$, the phase diagram contains two paramagnetic regions, two critical points, and one ferromagnetic region.

We adopt the quantum quench protocol established in Ref.~\cite{Calabrese:2005in}. The initial state $|\psi(0)\rag$ is prepared as an eigenstate of the Ising chain with Hamiltonian $H(h_0)$, where $h_0$ denotes the pre-quench transverse field. The state is then evolved unitarily under the post-quench Hamiltonian $H(h)$ as
\be
|\psi(t)\rag = \ep^{-\ii H(h)t} |\psi(0)\rag.
\ee
Following Refs.~\cite{Alba:2009th,Bucciantini:2014urw}, we work in the thermodynamic limit $L\to\inf$ and consider general eigenstates of the pre-quench Hamiltonian, each characterized by a function $m(\vph)\in[0,1]$ for momenta $\vph\in(-\pi,\pi]$. The function $m(\vph)$ can be interpreted as the excitation fraction of the mode at momentum $\vph$. The initial states we consider are Gaussian states, fully characterized by the two-point functions of Majorana modes \cite{Vidal:2002rm,Latorre:2003kg,Alba:2009th}, and the post-quench states $|\psi(t)\rag$ remain Gaussian at all times \cite{Calabrese:2005in,Bucciantini:2014urw}.

For a finite interval $A=[0,\ell]$ embedded in an infinite system with $L\to+\inf$, the RDM $\rho_A(t)=\tr_{\bar A}|\psi(t)\rag\lag\psi(t)|$ relaxes to the RDM of the GGE $\rho_{A,\GGE}=\tr_{\bar A}\rho_\GGE$ in the long-time limit $t \to +\inf$. This relaxation behavior holds regardless of whether the initial state is the ground state \cite{Calabrese:2011vdk,Calabrese:2012fmm,Calabrese:2012lgi} or an excited eigenstate \cite{Bucciantini:2014urw} of the pre-quench Hamiltonian.

A core open question is how $\rho_A(t)$ approaches $\rho_{A,\GGE}$ during the relaxation process. Ref.~\cite{Fagotti:2013jzu} addressed this by computing a distance measure $D(\rho_A(t),\rho_{A,\GGE})$ between the two RDMs. However, that work has three limitations. First, it only considers initial states corresponding to the ground state of the pre-quench Hamiltonian. Second, it does not survey the full parameter space of quench parameters $(h_0,h)$. Third, the distance measure adopted there, defined as
\be \label{ND}
D(\r,\s) = \f{\tr(\r-\s)^2}{\sqrt{\tr\r^2+\tr\s^2}},
\ee
has not been proven to be a mathematically rigorous distance metric. In this work, we aim to overcome all these limitations.

We use the recently proposed algorithm in \cite{Guo:2025jsm} and numerically compute the Bures distance $B_A(t) = B(\r_A(t),\r_{A,\GGE})$ between the time-dependent RDM $\rho_A(t)$ and the GGE RDM $\rho_{A,\GGE}$.
For finite $\ell$ in the late-time limit, we find that the subsystem distance follows a universal power-law decay
\be
B_A(t) \sim t^{-\lam},
\ee
where the decay exponent $\lam$ takes discrete values $1$, $5/4$, $3/2$, $7/4$, $2$, $5/2$, and possibly other values. The specific value of $\lam$ depends on the pre-quench and post-quench transverse fields $(h_0,h)$, as well as the function $m(\vph)$ that defines the initial eigenstate of $H(h_0)$. We systematically explore a broad range of $(h_0,h)$ values and various forms of $m(\vph)$. The result $\l=3/2$ reported in \cite{Fagotti:2013jzu} corresponds to a special case of the more general findings we present below.

\section{Typical transverse fields} \label{sec_typ}

\begin{figure}[t]
  \centering
  \includegraphics[width=0.45\textwidth]{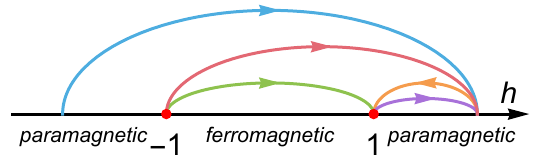}\\
  \caption{Five representative sets of pre-quench and post-quench transverse fields $(h_0,h)$: $(-2,2)$, $(2,1)$, $(1,2)$, $(-1,1)$, and $(-1,2)$, which collectively cover all qualitatively distinct parameter regimes.}
  \label{figh0h}
\end{figure}

By scanning a wide range of pre-quench and post-quench transverse field pairs $(h_0,h)$, we find that the following five typical sets suffice to represent all qualitatively distinct cases
\be
(h_0,h)=(-2,2),(2,1),(1,2),(-1,1),(-1,2).
\ee
These five representative parameter sets are illustrated in Figure~\ref{figh0h}.
\begin{itemize}
  \item $(-2,2)$: This set covers all cases where neither $h_0$ nor $h$ lies at a critical point, regardless of whether the two fields belong to the same or different ferromagnetic or paramagnetic phases. Examples include $(-2,2)$, $(3,2)$, $(-0.2,0.5)$, $(0.5,2)$, and $(2,0.5)$.
  \item $(2,1)$: This set covers all cases where $h_0$ is non-critical while $h$ is critical, regardless of whether $h_0$ falls in the ferromagnetic or paramagnetic phase, and regardless of whether the non-critical region is adjacent to the critical point. Examples include $(2,1)$, $(0.5,1)$, and $(-2,1)$.
  \item $(1,2)$: This set covers all cases where $h_0$ is critical while $h$ is non-critical, with the non-critical region adjacent to the critical point. Examples include $(1,2)$ and $(1,0.5)$.
  \item $(-1,1)$: This set corresponds to cases where both $h_0$ and $h$ are critical, covering the pairs $(1,-1)$ and $(-1,1)$.
  \item $(-1,2)$: This set covers all cases where $h_0$ is critical while $h$ lies in the paramagnetic phase, with the paramagnetic region not adjacent to the critical point. Examples include $(-1,2)$ and $(1,-2)$.
\end{itemize}

\section{Identifying discrete decay exponents via $m_S(\varphi)$} \label{sec_ide}

For any fixed pair of pre- and post-quench transverse fields $(h_0,h)$ falling into one of the five representative cases outlined above, the decay exponent depends on $m(\varphi)$ solely through the properties of the function \cite{Alba:2009th,Bucciantini:2014urw}
\be
m_S(\varphi) = m(\varphi)+m(-\varphi)-1.
\ee
This function is even and $2\pi$-periodic, satisfying $m_S(\varphi)=m_S(-\varphi)=m_S(\varphi+2\pi)$, which allows us to restrict its domain to $\varphi\in[0,\pi]$.

\subsection{States with discontinuous $m_S(\varphi)$}

When the function $m_S(\varphi)$ for $\varphi \in [0,\pi]$ is discontinuous, the decay exponent of the subsystem distance always takes the value $1$, independent of the pre- and post-quench transverse fields $(h_0,h)$.

\subsection{States with continuous $m_S(\varphi)$ and finite $m'_S(\varphi)$}

Continuity of $m_S(\varphi)$ over $\varphi\in[0,\pi]$ does not imply continuity of its first derivative $m'_S(\varphi)$. In this subsection, we consider cases where $m_S(\varphi)$ is continuous and its first derivative $m'_S(\varphi)$ is finite for all $\vph\in[0,\pi]$.

In addition to the pre- and post-quench transverse fields $(h_0,h)$, the decay exponent $\lambda$ is determined by three properties of $m_S(\varphi)$: whether the function vanishes at $\varphi=0$ and $\varphi=\pi$, whether it is smooth at these two boundaries, and whether it is smooth at all interior points $\varphi\in(0,\pi)$. Throughout this work, smoothness refers to $C^1$ continuity, meaning both $m_S(\varphi)$ and its first derivative $m'_S(\varphi)$ are continuous. By the even symmetry of $m_S(\varphi)$, $C^1$ continuity at $\varphi=0$ implies $m'_S(0)=0$; the same condition holds at $\varphi=\pi$ due to periodicity and reflection symmetry. For interior points $\varphi\in(0,\pi)$, $C^1$ continuity reduces to continuity of $m'_S(\varphi)$.

\begin{figure}[t]
  \centering
  \includegraphics[height=0.4\textwidth]{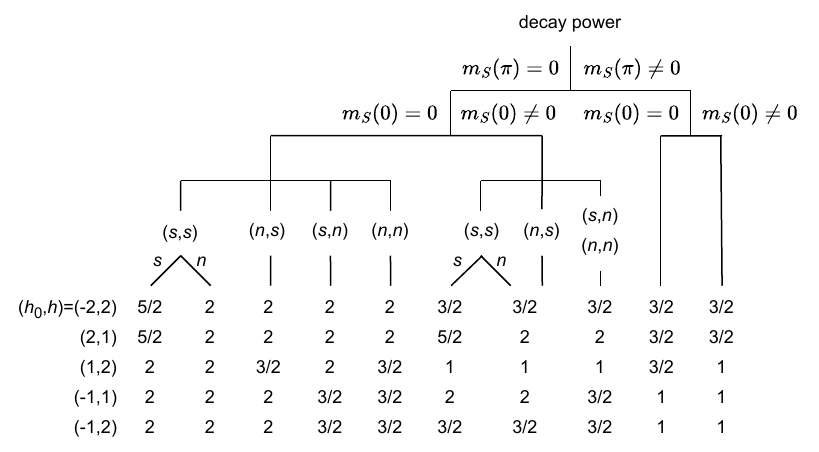}
  \caption{Flowchart for identifying discrete decay exponents for states with continuous $m_S(\varphi)$ and finite $m'_S(\varphi)$. The label $(s/n,s/n)$ encodes the smoothness of $m_S(\vph)$ at the boundary points $\vph=0$ and $\vph=\pi$, respectively. The single labels $s$ and $n$ denote global smoothness and the presence of non-smooth points, respectively, for the interior interval $\vph\in(0,\pi)$.}
  \label{fig_cla_1}
\end{figure}

A flowchart for identifying the discrete decay exponents is presented in Figure~\ref{fig_cla_1}. To determine the decay exponent, we first check whether $m_S(0)$ and $m_S(\pi)$ vanish. For cases with $m_S(\pi)\neq 0$, the decay exponent depends on both whether $m_S(0)$ vanishes and the transverse field pair $(h_0,h)$. Note that for the initial ground state, $m(\vph)=0$ identically, from which it follows directly that $m_S(\vph)=-1$.

Otherwise, one must examine the smoothness of $m_S(\varphi)$ at $\vph=0$ and $\vph=\pi$, i.e. whether $m'_S(0)$ and $m'_S(\pi)$ vanish, which we encode using the notation
\be
(s/n,s/n),
\ee
where $s$ denotes smooth and $n$ denotes non-smooth. For two special cases, specifically the $(s,s)$ configuration with $m_S(0)=m_S(\pi)=0$, and the $(s,s)$ configuration with $m_S(0)\neq0$ and $m_S(\pi)=0$, an additional check is required: the smoothness of $m_S(\varphi)$ across the interior interval $\vph\in(0,\pi)$. We use a single label $s$ to indicate that $m_S(\varphi)$ is smooth at all points in $\vph\in(0,\pi)$, and $n$ to indicate that at least one non-smooth point exists within the interval $\vph\in(0,\pi)$.

\begin{figure}[t]
  \centering
  \includegraphics[width=\textwidth]{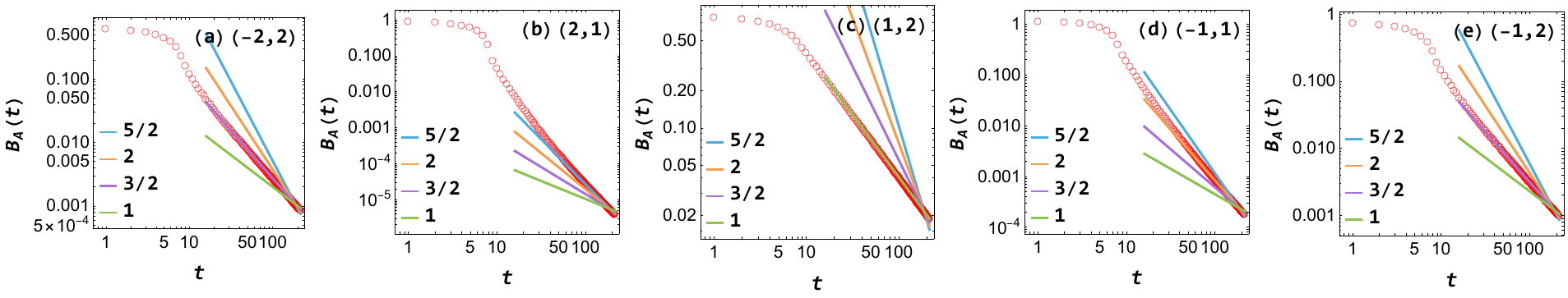}\\
  \caption{Decay of the subsystem Bures distance (empty red circles) following a global quench in the infinite Ising chain. Panels from left to right correspond to quench parameter sets $(h_0,h)=(-2,2)$, $(2,1)$, $(1,2)$, $(-1,1)$, and $(-1,2)$, respectively. Solid lines denote power-law fits with discrete decay exponents $1$, $3/2$, $2$, and $5/2$. The initial state is $m(\vph)=\f12 \sin^2\f{\vph}{2}$, which indicates $m_S(\vph)=-\cos^2\f{\vph}{2}$, and the subsystem length is set to $\ell=16$.}
  \label{fig_dec_pow}
\end{figure}

Examples of the subsystem Bures distance following a global quench in the infinite Ising chain are presented in Figure~\ref{fig_dec_pow}. The initial state is taken as $m(\vph)=\f12 \sin^2\f{\vph}{2}$. According to the classification flowchart in Figure~\ref{fig_cla_1}, the decay exponents corresponding to the parameter sets $(h_0,h)=(-2,2),(2,1),(1,2),(-1,1),(-1,2)$ are $\f32,\f52,1,2,\f32$, respectively.

\subsection{States with continuous $m_S(\varphi)$ and infinite $m'_S(\varphi)$}

In this subsection, we consider states for which $m_S(\varphi)$ is continuous over $\varphi\in[0,\pi]$, while its first derivative $m'_S(\varphi)$ diverges at one or more points in the interval. In such cases, the decay exponent may depend on the odd antisymmetric function $m_A(\varphi) \equiv m(\varphi)-m(-\varphi)$. However, this dependence is overly involved and falls outside the scope of this work; we therefore restrict our analysis to the special case $m_A(\varphi)=0$.

\begin{figure}[t]
  \centering
  \includegraphics[height=0.4\textwidth]{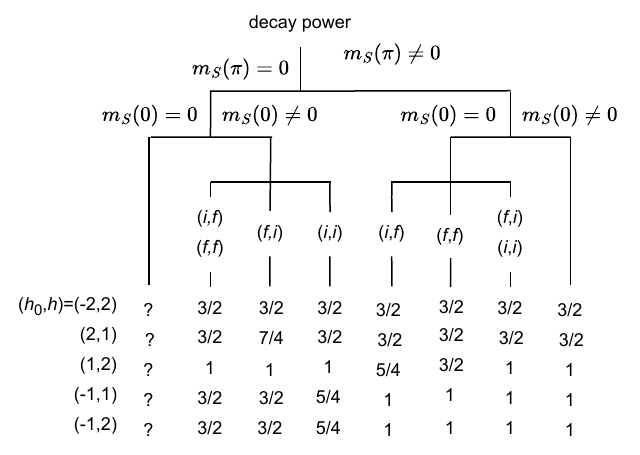}
  \caption{Flowchart for identifying discrete decay exponents for states with continuous $m_S(\varphi)$, divergent $m'_S(\varphi)$, and vanishing $m_A(\varphi)$. The label $(f/i,f/i)$ encodes whether $m'_S(\varphi)$ is finite ($f$) or divergent ($i$) at the boundary points $\varphi=0$ and $\varphi=\pi$, respectively. The branch corresponding to $m_S(0)=m_S(\pi)=0$ is left unspecified.}
  \label{fig_cla_2}
\end{figure}

Besides the pre- and post-quench transverse fields $(h_0,h)$, the decay exponent is governed by at least two properties of $m_S(\varphi)$: whether the function vanishes at $\varphi=0$ and $\varphi=\pi$, and whether $m'_S(\varphi)$ diverges at these two boundary points. Furthermore, for cases with $m_S(0)=m_S(\pi)=0$, the exponent also depends on whether $m'_S(\varphi)$ remains finite at all interior points $\varphi\in(0,\pi)$, as well as other possible properties of $m_S(\varphi)$. A complete closed-form rule for determining the decay exponent has not yet been established. We therefore present a flowchart in Figure~\ref{fig_cla_2} to identify the discrete decay exponents for the remaining three cases.

To determine the exponent, we first check whether $m_S(0)$ and $m_S(\pi)$ vanish. For cases where $m_S(\pi) \neq 0$ and $m_S(0) \neq 0$, the decay exponent is set by the transverse field pair $(h_0,h)$. For the remaining cases with $m_S(\pi) \neq 0$ and $m_S(0) = 0$, we further examine the behavior of $m'_S(\varphi)$ at the boundaries $\varphi=0$ and $\varphi=\pi$, which we encode using the ordered pair notation
\be
(f/i,f/i),
\ee
where $f$ denotes finite and $i$ denotes infinite.

\section{Conclusion and discussion} \label{sec_con}

In this work, we numerically investigate the late-time decay of subsystem distance following a global quantum quench in the infinite one-dimensional transverse-field Ising chain. Using the mathematically rigorous Bures distance to quantify the discrepancy between the time-evolved RDM and its GGE stationary state, we show that relaxation universally obeys a power-law form $B_A(t) \sim t^{-\lambda}$, with the decay exponent $\lambda$ taking only discrete values: $1$, $5/4$, $3/2$, $7/4$, $2$, $5/2$, and potentially others. We establish a classification scheme: the specific value of $\lambda$ is jointly determined by the pre- and post-quench transverse fields and the analytic properties of the symmetric function $m_S(\varphi)$, including its continuity, boundary values at $\varphi=0,\pi$, the regularity of its first derivative, and other possible features. The previously reported $t^{-3/2}$ decay for ground-state initial conditions is naturally recovered as a special case of our general framework.

We verify the robustness of this discrete power-law rule across multiple distance measures. Besides the Bures distance, we examine the definition used in Ref.~\cite{Fagotti:2013jzu} [Eq.~(\ref{ND})] and the ``relative distance'' $R(\rho,\sigma)\equiv\sqrt{2S(\rho \| \sigma)}$ \cite{Zhang:2024qya}, constructed from the quantum relative entropy $S(\rho \| \sigma) \equiv \operatorname{tr}(\rho\log\rho) - \operatorname{tr}(\rho\log\sigma)$. The discrete exponent classification remains consistent across all these metrics, indicating that our findings are insensitive to the specific choice of distance measure.

It was demonstrated analytically and numerically that the deviation of the post-quench transverse magnetization from its stationary value also follows discrete power-law decay~\cite{Bucciantini:2014urw}
\begin{equation} \label{mzt}
| m_z(t) - m_z(+\infty) | \sim t^{-\tilde{\lambda}},
\end{equation}
with reported examples of exponents $\tilde{\lambda}=1, 3/2, 5/2$. From Hölder's inequality
\begin{equation}
|\operatorname{tr}[ \mathcal{\mathcal{O}} (\rho-\sigma) ]| \leq 2 s_\mathcal{\mathcal{O}} D(\rho,\sigma),
\end{equation}
where $\mathcal{\mathcal{O}}$ is a generic operator on the shared Hilbert space of $\rho$ and $\sigma$, and $s_\mathcal{\mathcal{O}}$ is its largest singular value, combined with the inequality relating the trace distance $D$ and Bures distance $B$
\begin{equation}
D \leq B \sqrt{ 1 - {B^2}/{4} },
\end{equation}
we derive the bound $\tilde{\lambda} \geq \lambda$ relating the magnetization exponent to the subsystem distance exponent. Our numerical tests on a wide range of states consistently yield $\tilde{\lambda}=\lambda$ in all cases studied. While the magnetization in Eq.~(\ref{mzt}) is computationally cheaper to evaluate, it exhibits stronger oscillations than the subsystem Bures distance $B_A(t)$ for moderately large subsystem sizes. Extracting the decay exponent from the subsystem distance is therefore numerically more reliable.

Although we have explored a broad set of initial states and quench parameter pairs, it is not feasible to exhaust all possible configurations. Our conclusions are based on the numerically studied examples, and we do not rule out that the rule may break down for certain special states, or that additional discrete exponents may exist. Since our results rest primarily on numerical fitting, a rigorous analytical derivation of the discrete decay exponents remains an important direction for future work. Finally, our analysis is restricted to the one-dimensional transverse-field Ising model; it would be highly interesting to examine whether discrete subsystem distance decay persists in other one-dimensional integrable models and in higher-dimensional systems.

It is also instructive to study the scaling of $B_A$ with subsystem size $\ell$ in the regime $t \gg \ell \gg 1$. Fitting the ansatz $B_A\sim \ell^\alpha / t^\lambda$, we find $\alpha=2$ for most examples, consistent with the ground-state result of Ref.~\cite{Fagotti:2013jzu}. We also identify various cases with $\alpha=1$, along with tentative evidence for $\alpha=3/2$ in some instances. The $\alpha=3/2$ scaling cannot be firmly confirmed, however, as numerical errors grow substantially at large $t$. We have not identified a simple criterion that determines the exponent $\alpha$. A promising avenue for future work is to numerically establish or analytically derive the value of $\alpha$ in generic settings.

\section*{Acknowledgements}

We acknowledge support from the National Natural Science Foundation of China (NSFC) grant number 12205217 and Tianjin University Self-Innovation Fund Extreme Basic Research Project grant number 2025XJ21-0007.
Numerical calculations for this study were performed at high performance cluster at Center for Joint Quantum Studies (HPC-CJQS) of Tianjin University.

\appendix


\providecommand{\href}[2]{#2}\begingroup\raggedright\endgroup

\end{document}